\documentclass[12pt]{iopart}

\usepackage{graphicx}
\def\prl{Phys. Rev. Lett.}
\def\prd{Phys. Rev. D}

\def\pr{Phys. Rev.}

\begin{document}
   
\title{A Simple Family of Analytical Trumpet Slices of the Schwarzschild Spacetime}

\author{Kenneth A. Dennison and Thomas W. Baumgarte}
\address{Department of Physics and Astronomy, Bowdoin College,
  Brunswick, ME 04011, USA}
\ead{kdenniso@bowdoin.edu, tbaumgar@bowdoin.edu}
   
\begin{abstract}
We describe a simple family of analytical coordinate systems for the Schwarzschild spacetime.  The coordinates penetrate the horizon smoothly and are spatially isotropic.  Spatial slices of constant coordinate time $t$ feature a trumpet geometry with an asymptotically cylindrical end inside the horizon at a prescribed areal radius $R_0$ (with $0<R_{0}\leq M$) that serves as the free parameter for the family.  The slices also have an asymptotically flat end at spatial infinity.  In the limit $R_{0}=0$ the spatial slices lose their trumpet geometry and become flat -- in this limit, our coordinates reduce to Painlev\'e-Gullstrand coordinates. 
\end{abstract}

\pacs{04.20.Jb, 04.70.Bw, 97.60.Lf, 04.25.dg}

\section{Introduction}
\label{Intro}

The Schwarzschild spacetime can be described analytically in many different coordinate systems.  In addition to the original Schwarzschild coordinates \cite{Sch16},  well-known coordinate systems include Kruskal-Szekeres coordinates \cite{Kru60,Sze60}, Eddington-Finkelstein \cite{Edd24,Fin58} (or Kerr-Schild \cite{KerS65}) coordinates, harmonic (or De Donder) coordinates \cite{Ded21} as well as Painlev\'e-Gullstrand coordinates \cite{Pai21,Gul22}.   Another example is a one-parameter family of analytical coordinate systems that has both Eddington-Finkelstein and Painlev\'e-Gullstrand coordinates as members \cite{Lak94,MarP01} (see also  \cite{GauH78,Gau95}).

In this short paper we present another family of analytical coordinate systems representing the Schwarzschild spacetime.  We believe that this family has some remarkable properties: the coordinates extend smoothly through the black hole event horizon, the spatial coordinates are isotropic (so that the spatial metric can be written as a conformal factor to some power times a flat spatial metric), and, for almost all members of the family, the spatial slices take a so-called {\em trumpet} geometry.  Moreover, all expressions are surprisingly simple, particularly for one special member of the family.  

\begin{figure}[t]
\centering
\includegraphics[width=3in]{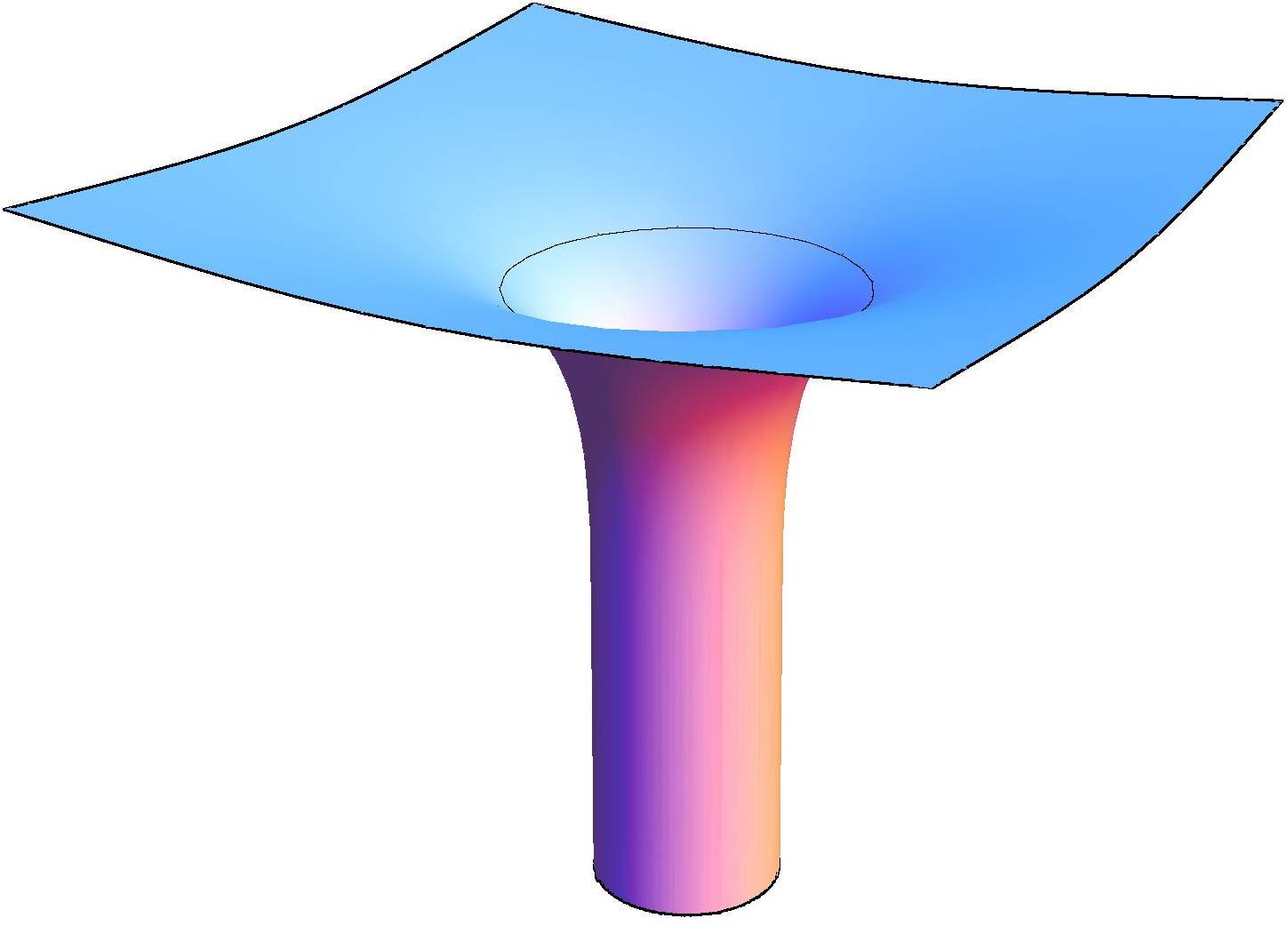}
\caption{Embedding diagram for the $R_{0}=M$ member of our family of solutions with $t=\rm{constant}$, $\theta=\pi/2$.  The distance from the axis of symmetry measures the areal radius $R$.  The circle near the top of the figure marks the event horizon at $R=2M$.}
\label{Fig1}
\end{figure}

Trumpet slices, meaning spatial slices of constant coordinate time that feature a trumpet geometry, have played an important role in numerical relativity, since they help numerical simulations avoid the spacetime singularities at the centers of black holes (see, e.g., \cite{HanHOBO08,BauS10} for discussions).  A trumpet slice ends on a sphere of non-zero (and finite) areal radius.   The proper distance between this sphere and any point away from the sphere, measured on a slice of constant coordinate time, is infinite.  Represented in an embedding diagram (see Fig.~\ref{Fig1}, or Fig.~2 in \cite{HanHOBO08}), the slice therefore appears to approach a cylinder.  The resulting shape has given the trumpet geometry its name.  Represented in a Penrose diagram, trumpet slices connect spatial infinity in one universe with future timelike infinity in the other universe (see Fig. \ref{Fig2} below). 

In numerical relativity simulations, trumpet slices emerge as a result of the imposed slicing condition.  In particular, the so-called 1+log slicing \cite{BonMSS95} leads, at late times, to ``stationary 1+log" trumpet slices \cite{HanHPBO06,HanHOBO08}.   A ``non-advective" version of the 1+log slicing leads to maximally sliced trumpet slices \cite{HanHOBGS06}.   While the latter can be expressed analytically, albeit only in parametric form \cite{BauN07}, it does not appear to be possible to express the former completely analytically.  Here we present a completely analytical family of trumpet slices.  The family is parameterized by the areal radius of the trumpet, $R_0$, and takes a particularly simple form for $R_0 = M$.  At the other limit of the family, $R_0 = 0$, we recover Painlev\'e-Gullstrand coordinates, for which the trumpet geometry disappears. 

This paper is organized as follows: In Section \ref{Sol} we present the family of solutions.  We follow this with a derivation of the family in Section \ref{Deriv}.  In Section \ref{NumRel} we discuss our solutions from the perspective of numerical relativity.  We conclude with a brief summary in Section \ref{Sum}.

%======================================================================
\section{A family of isotropic trumpet slices of the Schwarzschild spacetime}
\label{Sol}
%======================================================================

Consider the line element 
\begin{equation} \label{lineel}
ds^{2} = -\frac{r + R_0 - 2M}{r + R_0} \,dt^{2} + \frac{2f_{1}}{r} \, dtdr + \left(1+\frac{R_{0}}{r}\right)^{2}\left(dr^{2} + r^{2}d\Omega^{2}\right).
\end{equation}
Here we have used spherical polar coordinates with an isotropic radius $r$, we have abbreviated
\begin{equation} \label{f1def}
f_{1}\left(r\right)\equiv\sqrt{2r\left(M-R_{0}\right)+R_{0}\left(2M-R_{0}\right)},
\end{equation}
and $M$ and $R_0$ are constants.  It can be verified that the line element (\ref{lineel}) satisfies Einstein's equations in vacuum, but we will also derive this form of the metric from the Schwarzschild solution below.   In the following we restrict our analysis to $R_0 \leq M$ so that $f_1$ remains real for all $r \geq 0$.

Computing the proper area of a sphere centered on the origin we see that the areal radius $R$ is related to the isotropic radius $r$ by the simple relation
\begin{equation} \label{areal_isotropic}
R = r + R_0.
\end{equation}
In particular, this implies that the point $r=0$ corresponds to a sphere of areal radius $R_0$.  We also see that, for positive $R_0$, the proper distance from $r=0$ to any point $r > 0$ (at constant coordinate time $t$) is infinite.  Together, these two properties establish the spatial geometry of the line element (\ref{lineel}) as a trumpet geometry.  An embedding diagram of this geometry is shown in Fig.~\ref{Fig1}. 

\begin{figure}[t]
\label{Fig2}
\centering
\begin{minipage}{0.3\textwidth}
\includegraphics[width=\textwidth]{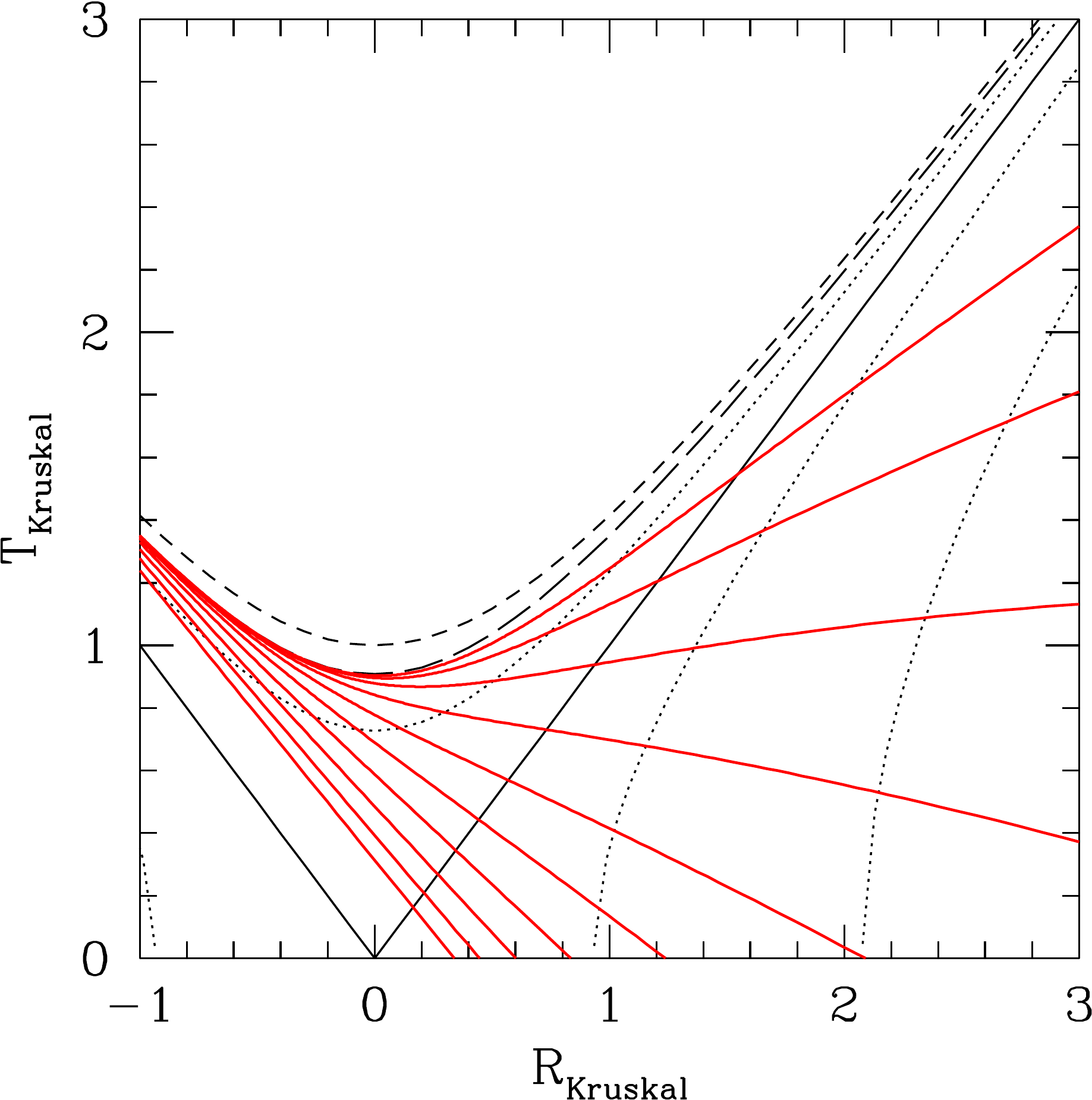}
\end{minipage}
\quad
\begin{minipage}{0.5\textwidth}
\includegraphics[width=\textwidth]{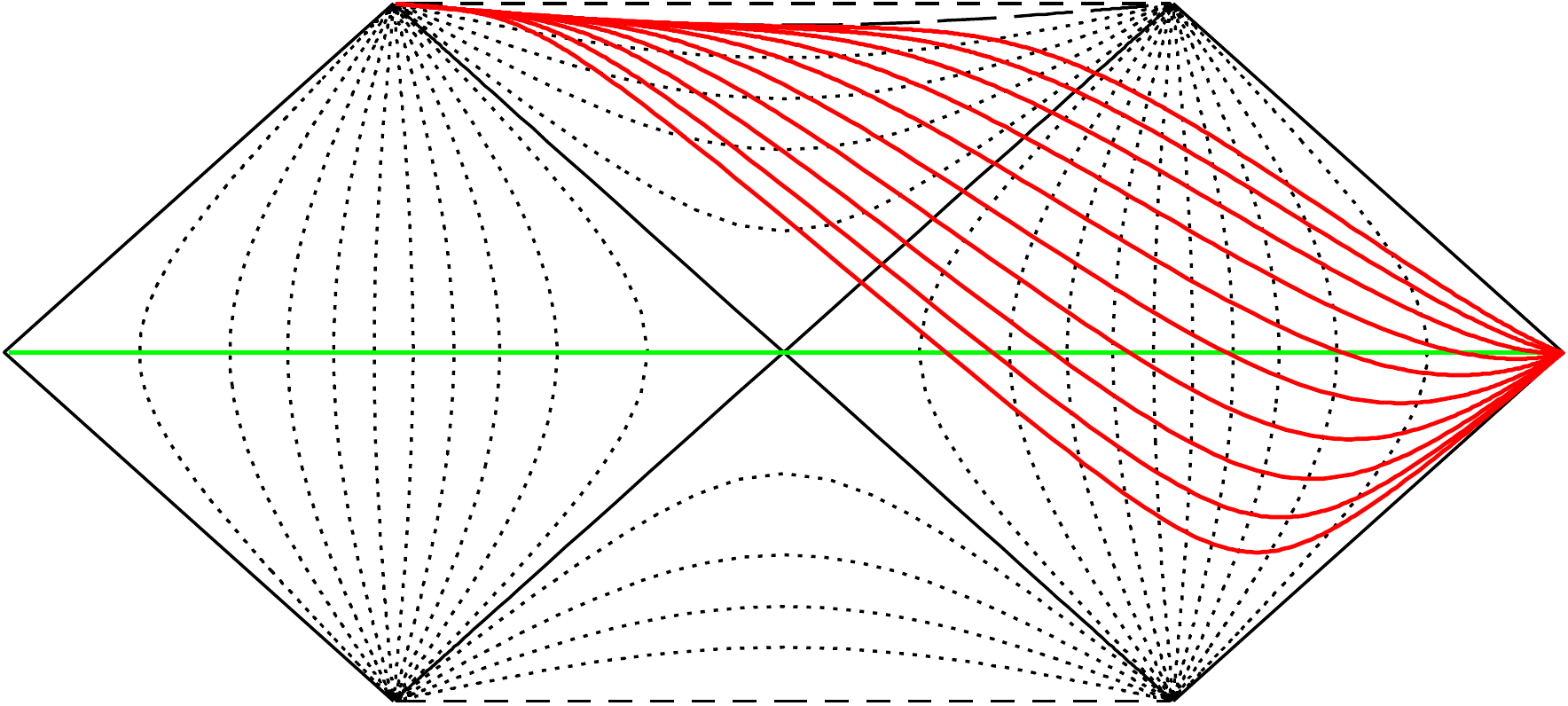}
\end{minipage}
\caption{Kruskal (left) and Penrose (right) diagrams for the $R_{0}=M$ member of the family of solutions (\ref{lineel}).  Short dashes mark the singularity at $R=0$, long dashes mark the limiting surface at $R=M$, and dots mark more general curves of constant areal radius $R$.  Solid curves (red online) mark $t=\rm{constant}$ trumpet slices which connect spatial infinity to future timeline infinity.  For comparison, the solid horizontal line (green online) in the Penrose diagram marks a wormhole slice.  Finally, the solid diagonal lines mark the event horizon.}
\end{figure}

Particular values of $R_{0}$ result in very simple solutions.  Letting $R_{0}=M$, we see that $f_{1}=M$, and the line element (\ref{lineel}) reduces to
\begin{equation}
\label{lineelR0M}
ds^{2} = -\frac{r-M}{r+M} \,dt^{2} + \frac{2M}{r} \,dtdr + \left(1+\frac{M}{r}\right)^{2}\left(dr^{2} + r^{2}d\Omega^{2}\right). 
\end{equation}
The relation (\ref{areal_isotropic}) now becomes $R = r + M$, which is the same relation as for harmonic (or De Donder) coordinates \cite{Ded21}.  Figure \ref{Fig2} shows Kruskal and Penrose diagrams for this solution.  If we choose $R_{0}=0$ instead, the line element (\ref{lineel}) becomes
\begin{equation}
\label{lineelPG}
ds^{2} = -\left(1-\frac{2M}{r}\right)dt^{2} + 2\sqrt{\frac{2M}{r}}\,dtdr + dr^{2} + r^{2}d\Omega^{2},
\end{equation}
which is well known as the Painlev\'e-Gullstrand line element.  

%============================================================================
\section{Transformation from Schwarzschild coordinates}
\label{Deriv}
%============================================================================

A straightforward derivation of the line element (\ref{lineel}) starts with the Schwarzschild solution in Schwarzschild coordinates, 
\begin{equation} \label{Schwarzschildle}
ds^{2} =-f_{0} \,d\bar{t}^{2} + f_{0}^{-1} \,dR^{2}+R^{2}d\Omega^{2}.
\end{equation}
Here $f_{0}\left(R\right)\equiv 1-2M/R$, and $M$ is the black hole's gravitational mass.   We then introduce a height function $h(R)$ that transforms the Schwarzschild time $\bar t$ to a new time coordinate 
\begin{equation}
\label{deftheight}
t = \bar{t} + h(R).
\end{equation}
In terms of this new time coordinate the line element takes the form 
\begin{equation}
\label{Schwarzschildleh}
ds^{2} = -f_{0}\,dt^{2} + 2f_{0}\frac{dh}{dR}\,dtdR + \left(f_{0}^{-1}-f_{0}\left(\frac{dh}{dR}\right)^{2}\right)dR^{2} + R^{2}d\Omega^{2}.
\end{equation}
We seek transformations that bring the spatial part of the line element into isotropic form, meaning that we can write this spatial part as some overall factor times the flat metric.  Following convention we express the overall factor as the fourth power of a conformal factor $\psi$ and identify
\begin{equation}
\label{lineelementspatialparts}
\left(f_{0}^{-1}-f_{0}\left(\frac{dh}{dR}\right)^{2}\right)dR^{2}+R^{2}d\Omega^{2}=\psi^{4}\left(dr^{2}+r^{2}d\Omega^{2}\right),
\end{equation}
where $r$ is again an isotropic radius.  From the angular part of this identification we obtain
\begin{equation} \label{findpsi}
\psi = \sqrt{\frac{R\left(r\right)}{r}},
\end{equation}
while the radial part yields
\begin{equation}
\label{finddhdR}
\frac{dh}{dR} = \frac{1}{f_{0}}\sqrt{1-f_{0}\left(\frac{R\left(r\right)}{rR'(r)}\right)^{2}}.
\end{equation}
Here we interpret $R = R(r)$ as a function of $r$, and abbreviate $R'(r)\equiv dR/dr$.  Finally, in order to obtain trumpet solutions, we look for solutions for which the conformal factor scales with $r^{-1/2}$ for small $r$.  A surprisingly simple solution of this form results from the choice
\begin{equation}
\label{defRr}
R\left(r\right) = r + R_{0},
\end{equation}
with $0<R_{0} \leq M$.  We then see that
\begin{equation}
\label{findpsieval}
\psi = \sqrt{1+\frac{R_{0}}{r}},
\end{equation}
which, for small $r$, diverges with $r^{-1/2}$ as desired.  Substituting equation (\ref{defRr}) into (\ref{finddhdR}) yields $dh/dR = f_1/(r f_0)$; inserting this into (\ref{Schwarzschildleh}) and expressing $f_0$ in terms of $r$ then results in the line element (\ref{lineel}) and completes the derivation.

%==============================================
\section{3+1 Decomposition}
\label{NumRel}
%==============================================

Since trumpet slices play a special role in numerical relativity, it is of interest to express the line element (\ref{lineel}) in terms of a $3+1$ decomposition (see, e.g., \cite{BauS10} for a textbook treatment).   Comparing the line element (\ref{lineel}) with the 3+1 form
\begin{equation} \label{3plus1lineelement}
ds^{2} = - \alpha^{2} dt^{2} + \gamma_{ij} \left(dx^{i} + \beta^{i} dt\right) \left(dx^{j} + \beta^{j} dt\right)
\end{equation}
we can identify the lapse function $\alpha$, the radial component of the shift vector $\beta^r$, and the spatial metric $\gamma_{ij}$ as
\begin{equation}
\label{lapse}
\alpha = \frac{r}{r + R_0},
~~~
\beta^{r} = \frac{rf_{1}}{\left(r + R_{0}\right)^{2}},
~~~
{\rm and}
~~~
\gamma_{ij} = \psi^{4}\eta_{ij}.
\end{equation}
Here the conformal factor $\psi$ is given by (\ref{findpsieval}), $\eta_{ij}$ is the flat metric in spherical polar coordinates, and the non-radial components of the shift vanish.  For time-independent   solutions, the extrinsic curvature can be computed from $K_{ij}=\left(D_{i}\beta_{j}+D_{j}\beta_{i}\right)/(2\alpha)$, where $D_i$ is the covariant derivative associated with $\gamma_{ij}$.   For the line element (\ref{lineel}) we find the non-zero components
\begin{equation}
\label{Krr}
K_{rr} = -\frac{r\left(M-R_{0}\right)+MR_{0}}{r^{2}f_{1}}
~~~{\rm and}~~~
K_{\theta\theta} = \frac{K_{\phi\phi}}{\sin^{2}\theta} = f_1,
\end{equation}
as well as the trace
\begin{equation} \label{K}
K \equiv \gamma^{ij} K_{ij} = \frac{\left(3r+2R_{0}\right)\left(M-R_{0}\right)+MR_{0}}{\left(r+R_{0}\right)^{2}f_{1}}.
\end{equation}
In many applications (for example in the BSSN formalism \cite{NakOK87,ShiN95,BauS98}) the extrinsic curvature is decomposed according to 
\begin{equation}
\label{Aij}
\tilde{A}_{ij}=\psi^{-4}\left(K_{ij}-\frac{1}{3}\gamma_{ij}K\right).
\end{equation}
All singular terms are then absorbed in the conformal factor, leaving the regular terms
\begin{equation}
\label{Arr}
\tilde{A}_{rr}=-\frac{2}{3}\frac{\left(3r+R_{0}\right)\left(M-R_{0}\right)+2 M R_{0}}{\left(r+R_{0}\right)^{2}f_{1}},
\end{equation}
and
\begin{equation}
\label{Athetatheta}
\tilde{A}_{\theta\theta}=\frac{\tilde{A}_{\phi\phi}}{\sin^{2}\theta} =\frac{r^{2}\left(r\left(M-R_{0}\right)+R_{0}\left(M- R_{0}/3\right)\right)}{\left(r+R_{0}\right)^{2}f_{1}}.
\end{equation}
For $R_0=M$ most of the above expressions simplify significantly. 

The 1+log slicing condition \cite{BonMSS95}, which has proven to be extremely valuable in numerical simulations of black holes, is a member of the family of slicing conditions
\begin{equation} \label{slicingcondition}
\left(\partial_{t}-\beta^{i}\partial_{i}\right)\alpha=-\alpha^{2}f\left(\alpha\right)K
\end{equation}
for the special choice $f(\alpha) = 2/\alpha$.  From the above expressions we see that the line element (\ref{lineel}) satisfies the slicing condition (\ref{slicingcondition}) if instead we choose
\begin{equation} \label{fofalpha}
f\left(\alpha\right) = \frac{1-\alpha}{\alpha} \, \frac{2M-R_{0}\left(1+\alpha\right)}{3M-R_{0}\left(2+\alpha\right)},
\end{equation}
or just $f(\alpha) = (1 - \alpha)/\alpha$ for $R_0 = M$.  Unfortunately, this does not appear to be a very promising choice from the perspective of numerical relativity.  As discussed in \cite{BonMSS95}, the properties of the resulting gauge speeds suggest that one should choose $f(\alpha) >1$; here, however, $f(\alpha) \rightarrow 0$ as $r \rightarrow \infty$.  We therefore do not expect the family of solutions (\ref{lineel}) to be of great practical use in numerical relativity, at least for this slicing condition. 

%================================================================
\section{Summary}
\label{Sum}
%================================================================

We present a one-parameter family of analytical coordinate representations of the Schwarzschild spacetime.  We believe that this family has some remarkable properties, in addition to being surprisingly simple: the coordinates penetrate smoothly through the event horizon, the spatial coordinates are isotropic, and the spatial slices feature a trumpet geometry.  The family is parameterized by the areal radius $R_0$ of the sphere to which the trumpets asymptote; for $R_0 = 0$ we recover Painlev\'e-Gullstrand coordinates.   While these coordinates may not be of great practical use in numerical relativity, we believe that they are interesting in their own right, and that they provide a simple pedagogical example of black holes in trumpet geometries.

\ack

We would like to thank Eric Gourgoulhon, Edward Malec, Eric Poisson, and Stu Shapiro for helpful conversations and comments.  This work was supported in part by NSF grant PHYS-1063240 to Bowdoin College.


\section*{References}
\begin{thebibliography}{10}

\bibitem{Sch16}
K.~{Schwarzschild}.
\newblock {\"Uber das Gravitationsfeld eines Massenpunktes nach der
  Einsteinschen Theorie}.
\newblock {\em Sitzber. Deut. Akad. Wiss. Berlin, Kl. Math.-Phys. Tech.}, pages
  189--196, 1916.

\bibitem{Kru60}
M.~D. {Kruskal}.
\newblock {Maximal Extension of Schwarzschild Metric}.
\newblock {\em \pr}, 119:1743--1745, 1960.

\bibitem{Sze60}
G.~{Szekeres}.
\newblock {On the Singularities of a Riemannian Manifold}.
\newblock {\em Publ. Mat. Debrecen.}, 7:285--301, 1960.

\bibitem{Edd24}
A.~S. {Eddington}.
\newblock {A Comparison of Whitehead's and Einstein's formul\ae}.
\newblock {\em Nature}, 113:192, 1924.

\bibitem{Fin58}
D.~{Finkelstein}.
\newblock {Past-Future Asymmetry of the Gravitational Field of a Point
  Particle}.
\newblock {\em Phys. Rev.}, 110:965--967, 1958.

\bibitem{KerS65}
R.~P. {Kerr} and A.~{Schild}.
\newblock {A new class of vacuum solutions of the Einstein field equations}.
\newblock In G.~{Barb\`{e}ra}, editor, {\em {Proceedings of the Galileo Galilei
  Centenary Meeting on General Relativity, Problems of Energy and Gravitational
  Waves}}, pages 222--233. {Comitato Nazionale per le Manifestazione
  Celebrative, Florence}, 1965.

\bibitem{Ded21}
T.~{De Donder}.
\newblock {\em La gravifique einsteinienne}.
\newblock Gauthier-Villars, Paris, 1921.

\bibitem{Pai21}
P.~{Painlev\'e}.
\newblock {La M\'ecanique classique et la th\'eorie de la relativit\'e}.
\newblock {\em {C. R. Acad. Sci. (Paris)}}, 173:677--680, 1921.

\bibitem{Gul22}
A.~{Gullstrand}.
\newblock {Allgemeine L\"osung des statischen Eink\"orper\-problems in der
  Einsteinschen Gravitations\-theorie}.
\newblock {\em {Arkiv. Mat. Astron. Fys.}}, 16(8):1--15, 1922.

\bibitem{Lak94}
K.~{Lake}.
\newblock {A class of quasi-stationary regular line elements for the
  Schwarzschild geometry}.
\newblock arXiv:gr-qc/9407005, 1994.

\bibitem{MarP01}
K.~{Martel} and E.~{Poisson}.
\newblock {Regular coordinate systems for Schwarzschild and other spherical
  spacetimes}.
\newblock {\em Am. J. Phys.}, 69:476--480, 2001.

\bibitem{GauH78}
R.~{Gautreau} and B.~{Hoffmann}.
\newblock {The Schwarzschild radial coordinate as a measure of proper
  distance}.
\newblock {\em \prd}, 17:2552--2555, 1978.

\bibitem{Gau95}
R.~{Gautreau}.
\newblock {Light cones inside the Schwarzschild radius}.
\newblock {\em {Am. J. Phys.}}, 63:431--439, 1995.

\bibitem{HanHOBO08}
M.~{Hannam}, S.~{Husa}, F.~{Ohme}, B.~{Br{\"u}gmann}, and N.~{\'O} {Murchadha}.
\newblock {Wormholes and trumpets: Schwarzschild spacetime for the
  moving-puncture generation}.
\newblock {\em \prd}, 78:064020/1--19, 2008.

\bibitem{BauS10}
T.~W. {Baumgarte} and S.~L. {Shapiro}.
\newblock {\em {Numerical Relativity: Solving Einstein's Equations on the
  Computer}}.
\newblock Cambridge University Press, Cambridge, England, 2010.

\bibitem{BonMSS95}
C.~{Bona}, J.~{Mass{\'o}}, E.~{Seidel}, and J.~{Stela}.
\newblock {New Formalism for Numerical Relativity}.
\newblock {\em \prl}, 75:600--603, 1995.

\bibitem{HanHPBO06}
M.~{Hannam}, S.~{Husa}, D.~{Pollney}, B.~{Br{\"u}gmann}, and N.~{\'O
  Murchadha}.
\newblock {Geometry and Regularity of Moving Punctures}.
\newblock {\em \prl}, 99:241102/1--4, 2007.

\bibitem{HanHOBGS06}
M.~{Hannam}, S.~{Husa}, N.~{\'O} {Murchadha}, B.~{Br{\"u}gmann}, J.~A.
  {Gonz{\'a}lez}, and U.~{Sperhake}.
\newblock {Where do moving punctures go?}
\newblock {\em {J.~Phys.~Conf.~Series}}, 66:012047/1--9, 2007.

\bibitem{BauN07}
T.~W. {Baumgarte} and S.~G. {Naculich}.
\newblock Analytical representation of a black hole puncture solution.
\newblock {\em \prd}, 75:067502/1--4, 2007.

\bibitem{NakOK87}
T.~{Nakamura}, K.~{Oohara}, and Y.~{Kojima}.
\newblock {General Relativistic Collapse to Black Holes and Gravitational Waves
  from Black Holes}.
\newblock {\em Prog. Theor. Phys. Suppl.}, 90:1--218, 1987.

\bibitem{ShiN95}
M.~{Shibata} and T.~{Nakamura}.
\newblock {Evolution of three-dimensional gravitational waves: Harmonic slicing
  case}.
\newblock {\em \prd}, 52:5428--5444, 1995.

\bibitem{BauS98}
T.~W. {Baumgarte} and S.~L. {Shapiro}.
\newblock {Numerical integration of Einstein's field equations}.
\newblock {\em \prd}, 59:024007/1--7, 1998.

\end{thebibliography}
\end{document}